\documentclass[conference,compsoc]{IEEEtran}

\usepackage{enumitem}
\setlist{leftmargin=5mm}  

\usepackage{titlesec}
\titlespacing\section{0pt}{10pt plus 2pt minus 1pt}{0pt plus 1pt minus 1pt}
\titlespacing\subsection{0pt}{10pt plus 2pt minus 1pt}{0pt plus 1pt minus 1pt}
\titlespacing\subsubsection{0pt}{10pt plus 2pt minus 1pt}{0pt plus 1pt minus 1pt}

\usepackage{hyperref}

\usepackage{array}
\usepackage{tabulary}
\usepackage{multirow}
\usepackage{listings}
\usepackage{caption}
\usepackage{subcaption}

\usepackage{comment}
\usepackage{todonotes}

\usepackage{graphicx}
\usepackage{xcolor}
\usepackage{xspace}
\usepackage{amsmath}
\usepackage{soul}

\usepackage{booktabs}

\usepackage{url}
\usepackage{breakurl}

\newcommand{\System}{\textsc{WAFFLED}\xspace}

\newcommand{\json}{\texttt{application/json}\xspace}
\newcommand{\xml}{\texttt{application/xml}\xspace}
\newcommand{\formdata}{\texttt{multipart/form-data}\xspace}
\newcommand{\normalizer}{\texttt{HTTP-Normalizer\xspace}\xspace}

\definecolor{codegreen}{rgb}{0,0.6,0}
\definecolor{codegray}{RGB}{184, 184, 184}
\definecolor{codepurple}{rgb}{0.58,0,0.82}
\definecolor{backcolour}{rgb}{0.95,0.95,0.92}
\definecolor{bg}{rgb}{0.95,0.95,0.95}
\definecolor{codelgreen}{RGB}{105, 120, 33}

\lstdefinestyle{mystyle}{
  backgroundcolor=\color{backcolour}, commentstyle=\color{codegreen},
  keywordstyle=\color{magenta},
  numberstyle=\tiny\color{codegray},
  stringstyle=\color{codepurple},
  basicstyle=\ttfamily\footnotesize,
  breakatwhitespace=false,         
  breaklines=true,                 
  captionpos=b,                    
  keepspaces=true,                 
  numbers=left,                    
  numbersep=5pt,                  
  showspaces=false,                
  showstringspaces=false,
  showtabs=false,                  
  tabsize=2
}

\lstdefinelanguage{HTTP}{
    morekeywords={POST,GET,Content-Length,Content-Type,Content-Disposition},
    sensitive=false,
    morecomment=[l]{--},
}
\lstdefinestyle{highlighted_http}{
    language=HTTP,
    basicstyle=\ttfamily\footnotesize,
    keywordstyle=\bfseries\color{codelgreen},
    commentstyle=\color{gray},
    showstringspaces=false,
    breaklines=true,
    frame=single,
    escapeinside={(*}{*)}, 
}

\makeatletter
\AtBeginDocument{\let\c@listing\c@lstlisting}
\makeatother

\begin{document}

\title{\System: Exploiting Parsing Discrepancies to Bypass Web Application Firewalls}

\author{\IEEEauthorblockN{1\textsuperscript{st} Seyed Ali Akhavani}
\IEEEauthorblockA{
\textit{Northeastern University}\\
Boston, MA, USA \\
sadatakhavani.s@northeastern.edu}
\and
\IEEEauthorblockN{2\textsuperscript{nd} Bahruz Jabiyev}
\IEEEauthorblockA{
\textit{Northeastern University}\\
Boston, MA, USA \\
jabiyev.bahruz@gmail.com}
\and
\hspace{-0.2cm}
\IEEEauthorblockN{3\textsuperscript{rd} Ben Kallus}
\IEEEauthorblockA{
\hspace{-0.2cm}
\textit{Dartmouth College}\\
\hspace{-0.1cm}Hanover, NH, USA \\
\hspace{-0.2cm}benjamin.p.kallus.gr@dartmouth.edu}
\and
\hspace{1.5cm}
\IEEEauthorblockN{4\textsuperscript{th} Cem Topcuoglu}
\IEEEauthorblockA{
\hspace{1.5cm}
\textit{Northeastern University}\\
\hspace{1.5cm}Boston, MA, USA \\
\hspace{1.5cm}topcuoglu.c@northeastern.edu}
\and
\hspace{1.0cm}
\IEEEauthorblockN{5\textsuperscript{th} Sergey Bratus}
\IEEEauthorblockA{
\hspace{1.0cm}
\textit{Dartmouth College}\\
\hspace{1.2cm}Hanover, NH, USA \\
\hspace{1.0cm}Sergey.L.Bratus@dartmouth.edu}
\and
\hspace{1.4cm}
\IEEEauthorblockN{6\textsuperscript{th} Engin Kirda}
\IEEEauthorblockA{
\hspace{1.4cm}
\textit{Northeastern University}\\
\hspace{1.4cm}Boston, MA, USA \\
\hspace{1.4cm}e.kirda@northeastern.edu}
}

\sloppy
\UseRawInputEncoding

\maketitle
\IEEEpeerreviewmaketitle

\begin{abstract}

Web Application Firewalls (WAFs) have been introduced as essential and popular security gates that inspect incoming HTTP traffic to filter out malicious requests and provide defenses against a diverse array of web-based threats. Evading WAFs can compromise these defenses, potentially harming Internet users. In recent years, parsing discrepancies have plagued many entities in the communication path; however, their potential impact on WAF evasion and request smuggling remains largely unexplored. In this work, we present an innovative approach to bypassing WAFs by uncovering and exploiting parsing discrepancies through advanced fuzzing techniques. By targeting non-malicious components such as headers and segments of the body and using widely used content-types such as \texttt{application/json}, \texttt{multipart/form-data}, and \texttt{application/xml}, we identified and confirmed 1207 bypasses across 5 well-known WAFs, AWS, Azure, Cloud Armor, Cloudflare, and ModSecurity. To validate our findings, we conducted a study in the wild, revealing that more than 90\% of websites accepted both \texttt{application/x-www-form-urlencoded} and \texttt{multipart/form-data} interchangeably, highlighting a significant vulnerability and the broad applicability of our bypass techniques. We have reported these vulnerabilities to the affected parties and received acknowledgments from all, as well as bug bounty rewards from some vendors. Further, to mitigate these vulnerabilities, we introduce \normalizer, a robust proxy tool designed to rigorously validate HTTP requests against current RFC standards. Our results demonstrate its effectiveness in normalizing or blocking all bypass attempts presented in this work.

\end{abstract}

\begin{IEEEkeywords}
Web Security, Web Application Firewalls, Fuzzing, HTTP, Parsing Discrepancies, Request Smuggling, WAF Evasion
\end{IEEEkeywords}

\section{Introduction}
\label{sec:intro}

The widespread adoption of web applications has made them prime targets for cyberattacks. To protect these applications, Web Application Firewalls (WAFs) have been introduced as essential and popular security gates. These systems inspect incoming HTTP traffic to filter out malicious requests, and provide defenses against a diverse array of web-based threats, ranging from SQL injection to Cross-Site Scripting attacks, and beyond.

Despite their critical role, WAFs are not immune to evasion. Traditional WAF evasion techniques often rely on distorting attack payloads to bypass detection rules while ensuring the payloads remain executable by web applications. Attackers usually either obfuscate the payload with encoding schemes or inject new characters into payloads to bypass WAF rules. However, WAF vendors have already taken measures against most of these bypass techniques, which have been known for many years now. Also, these types of attacks assume the ability of the target web application to parse the encoded, or obfuscated payload.

As the threat landscape evolves, HTTP Request Smuggling (HRS) attacks have gained significant attention. HRS exploits discrepancies in the interpretation of HTTP requests between different entities in the communication chain, such as servers, proxies, and WAFs. These attacks can have severe consequences, including unauthorized access to sensitive information, session hijacking, and server compromise. The increasing complexity of web applications and their reliance on intermediary components have increased the risk of HRS vulnerabilities. Recent studies have investigated the importance of HRS, identifying new variants of attack and suggesting defense mechanisms~\cite{grenfeldt2021attacking, portswigger2020hrs, blackhat2020hrs, klein2020hrs}.

Parsing discrepancies, caused by inconsistencies in the interpretation of HTTP requests, play a critical role in enabling attacks such as HRS. These discrepancies first appeared in the communication path between servers, but WAFs, which are integrated to this path, may themselves be vulnerable to such inconsistencies. As a layer between the client and the web application, WAFs must correctly interpret HTTP requests to protect against malicious activities. However, vulnerabilities in their parsing mechanisms can allow attackers to exploit these discrepancies, bypassing the WAF, and allowing attacks to reach the web application. Building on this understanding of the limitations of current WAF defenses and the emerging threat of HRS, we present a novel, real-world approach to bypassing WAFs. Our method exploits content parsing discrepancies between WAFs and web application frameworks. Unlike traditional evasion tactics, we keep the attack payload intact, and focus on mutating specific content elements, such as the boundary in \formdata, or namespace feature in \xml, causing the WAF to misinterpret the content. This misinterpretation allows the payload to pass through, while the web application framework correctly parses and executes the attack.

Our work tested a wide range of combinations of popular WAFs, including Google Cloud Armor, Cloudflare, AWS WAF, Azure WAF, and ModSecurity on NGINX, alongside widely-used web application frameworks such as Flask, Laravel, FastAPI, Gin, Express, and Spring Boot. We focused on three complex content types: \formdata, \xml, and \json. By repurposing a grammar-based and structure-aware HTTP fuzzer, we identified implementation differences in how these WAFs and frameworks parse these content types. Our findings reveal that most WAF-framework pairs can be bypassed using various content distortions, highlighting a significant vulnerability in current WAF implementations.

In this work, we summarize our contributions as follows:
\begin{itemize}
    \item We introduce a fundamentally new approach that uses content parsing discrepancies
    to bypass web application firewalls.
    \item We present a practical methodology for automatically finding new discrepancy-based bypass vectors using black-box fuzzing techniques.
    \item We design and implement a pipeline that tests these bypass vectors against popular web application firewalls and frameworks.
    \item We demonstrate successful discrepancy-based bypass instances on the pairs of most popular web application firewalls and frameworks, and we coordinate mitigation efforts with the affected technology vendors.
    \item We analyze the interchangeability of our bypass techniques using real-world data from PublicWWW, demonstrating that our findings are widely-applicable and practical across real-world web applications.
    \item We present the first tool, \normalizer, that ensures that all proposed bypasses are avoidable by enforcing proper techniques.
\end{itemize}

\textbf{Availability.} Our results and source code, including fuzzer input grammars and WAF configurations are publicly available\footnote{\url{https://github.com/sa-akhavani/waffled}}. Sensitive bypass requests will remain restricted until the vulnerabilities are resolved.

\section{Background}

\subsection{Web Content Types and RFCs Explained}
\label{appendix:contenttypes}

Our work focuses on using specific features of content-types to bypass a malicious request. Thus, it is crucial to inspect the details of the content and media types discussed in this paper.

\textbf{RFCs (Request for Comments).} 
RFCs~\cite{ietf_rfc_process} are a series of documents that define protocols, procedures, and conventions used on the Internet and networking standards. They are published by the Internet Engineering Task Force (IETF) and related organizations. Each RFC is assigned a unique number, and these documents serve as the authoritative source of information on various Internet standards and protocols. For instance, the structure of HTTP headers, MIME types, and various other web technologies are defined by respective RFCs.

\textbf{ABNF (Augmented Backus-Naur Form).} 
ABNF grammar rules~\cite{rfc5234_abnf} are a formal notation used to specify the syntax of RFCs. ABNF extends the basic BNF notation to provide a more flexible and precise way to define the syntax of Internet protocols. It is particularly important in our work as it allows for precise definitions of content types and headers, which we exploit to identify parsing discrepancies in WAFs. By understanding and manipulating ABNF rules, we can generate requests that challenge the ability of WAFs to consistently parse and enforce these standards.

\textbf{Multipart.} Multipart content-types allow HTTP messages to include multiple entities with varying media types in a single message body. They are crucial for handling file uploads and complex data structure support. These content-types, such as multipart/form-data, multipart/related, and multipart/mixed, are governed by specific RFCs (e.g., RFC 2387, RFC 7578, RFC 2045-2047)~\cite{masinter_returning_2015, levinson_mime_1998, freed_multipurpose_1996, freed_multipurpose_1996_p2, moore_mime_1996} that define their structure and usage scenarios. Our research investigates the complexities of multipart/form-data requests, exploiting features such as boundary definitions, charset, content-disposition, and other custom headers.
    
\textbf{XML.} Extensible Markup Language serves as a flexible format for structured data representation, commonly used in data exchange protocols. XML-based requests, defined in RFC 7303~\cite{thompson_xml_2014}, are characterized by elements such as DOCTYPE declarations, schemas, and CDATA sections. Our approach includes mutating these structural elements to investigate WAF responses to XML-specific parsing matters. For example, variations in DOCTYPE declarations and CDATA usage are tested to identify vulnerabilities in WAF handling of XML content.

\textbf{JSON.} JavaScript Object Notation, described in RFC 8259~\cite{bray_javascript_2017}, is a lightweight data interchange format widely used in modern web applications. JSON requests are characterized by their simple structure based on key-value pairs. We manipulate JSON object formatting and nested structures to assess WAF handling of JSON parsing anomalies.

\noindent\textbf{How WAFs Work.}
A WAF operates by inspecting HTTP traffic between clients and web applications, filtering out potentially harmful requests to prevent attacks such as  SQL injection, cross-site scripting (XSS), and malicious script uploads.
Positioned inline with the traffic flow, the WAF acts as a barrier between users and the application server, and is able to analyze requests in real-time before they reach the protected application.
To make filtering decisions, WAFs operate by scanning incoming HTTP requests, analyzing headers, cookies, URL parameters, and body content. Figure~\ref{fig:waf} illustrates this process. Numerous commercial and open-source WAFs implement these mechanisms, including Cloudflare, Google Cloud Armor, Microsoft Azure WAF, Amazon AWS WAF, and ModSecurity. WAFs typically rely on one or more detection strategies:

\begin{figure}[ht]
    \centering
    \includegraphics[width=0.9\columnwidth]{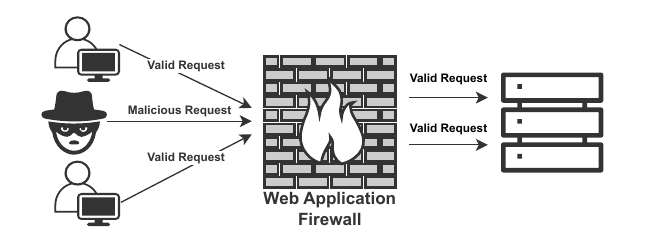}
    \caption{WAF filtering malicious requests.}
    \label{fig:waf}
    \vspace{-1em}
\end{figure}

\begin{figure*}[t!]
    \centering
    \includegraphics[width=0.75\linewidth]{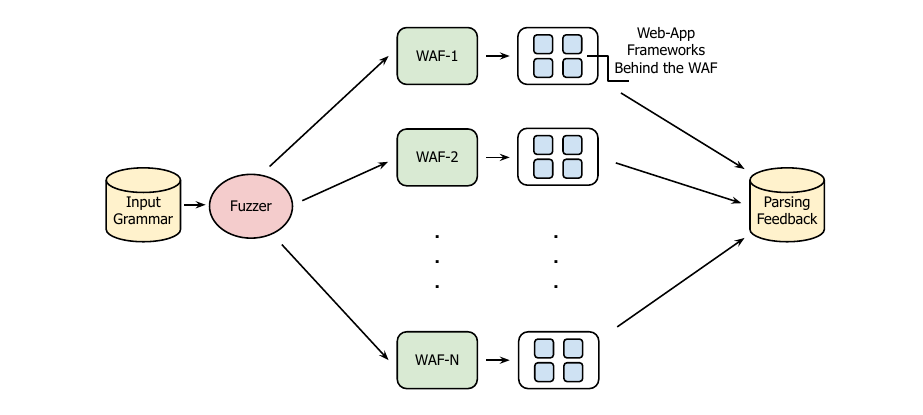}
    \caption{\System Overview.}
    \label{fig:diagram}
\end{figure*}

\textbf{Managed Rulesets.} WAFs use predefined sets of rules that define patterns associated with common attacks. These rulesets are frequently updated to respond to new threats. Most WAFs use their own custom rulesets. But one of the famous rulesets that is widely used among almost all WAFs and most custom rulesets use it as a base, is OWASP CRS (Core Rule Set)~\cite{owasp_crs}. This ruleset is a set of generic attack detection rules for use with compatible WAFs. It aims to protect web applications from a wide range of attacks, including the OWASP Top 10. CRS provides protection against many common attack categories, including SQL Injection, Cross Site Scripting, Local File Inclusion, etc. 

\textbf{Signature-Based Detection.} This method involves identifying known attack vectors based on specific signatures or patterns in the data. It is effective against known threats, but can be limited in identifying novel or modified attacks.

\textbf{Machine Learning-Based Detections.} 
Advanced WAFs incorporate Machine Learning models that analyze traffic behavior and adapt over time, improving their ability to detect previously unseen threats.

To accommodate complex attack patterns, WAF rules often rely on regular expressions (regex). For instance, detecting SQL injection attempts may involve scanning URL parameters or request bodies for SQL-related keywords. When a request arrives, the WAF extracts relevant fields, evaluates them against the defined rules, and executes the appropriate action. If a match is found, the WAF may block the request, log the event, or challenge the user with additional security measures such as a CAPTCHA. In cases where no match occurs, the request is forwarded to the application without interference. By using well-defined rules and regular expressions, a WAF effectively protects web applications from a wide range of attacks.

Each request is parsed to extract these fields, allowing the WAF to evaluate them against a predefined set of rules. These rules follow a structured format that are displayed in listing~\ref{listing:wafformat} and are as follows:

\vspace{0.5ex}
\lstset{style=mystyle}
    \begin{lstlisting}[language=Bash, frame=single, basicstyle=\small, label=listing:wafformat, caption={WAF Rule Format.}]
if <field><operator><value>
then <action>
\end{lstlisting}

\begin{itemize}
    \item \textbf{Field}: This could be any part of the HTTP request such as a header, cookie, MIME type, or URL parameter.
    \item \textbf{Operator}: Common operators include \textit{equals}, \textit{contains}, \textit{not equal}, etc.
    \item \textbf{Value}: The specific value to match against.
    \item \textbf{Action}: The action to be taken if the rule matches, such as \textit{block}, \textit{skip}, or \textit{send security challenge}.
\end{itemize}

\noindent\textbf{Example of a WAF in action.} Listing~\ref{listing:wafrule} illustrates a simple WAF rule definition and how it is applied during request inspection.

\lstset{style=mystyle}
\begin{lstlisting}[language=Bash, frame=single, basicstyle=\small, label=listing:wafrule, caption={Example WAF Rule.}]
if url_parameter "user_input" contains "DROP TABLE"
then block
\end{lstlisting}

In this case, when a request with the URL parameter \texttt{user\_input=DROP TABLE users} arrives, the WAF parses the request and extracts relevant fields. Then, it evaluates the value of \texttt{user\_input} against the defined rule. The pattern \texttt{"DROP TABLE"} is matched using a regular expression, triggering the \texttt{block} action. As a result, the WAF denies the request and logs the incident for further analysis.

While such rule-based filtering mechanisms are effective in many scenarios, a critical but often overlooked aspect of WAF behavior is how the system parses and interprets incoming data. Different WAFs may handle HTTP request parsing in different ways, particularly when dealing with outdated RFCs, nested data structures, or uncommon content types. These \textit{parsing discrepancies}, create opportunities for attackers to bypass protections by crafting inputs that are marked as valid by the WAF but would exploit the the target application. This work focuses on uncovering and categorizing such evasions by analyzing how parsing discrepancies arise and demonstrating practical techniques that abuse these mismatches to bypass modern WAFs.

\section{Related Work}
\label{sec:related_work}

At the architectural level, bypasses exploit vulnerabilities within the web application's infrastructure. Techniques such as IP spoofing and server-side request forgery (SSRF) allow attackers to access the origin server directly, circumventing WAF protection. Cache poisoning attacks, as discussed by Meiners et al.~\cite{cpdos}, exploit CDN vulnerabilities to disrupt access to web applications. Additionally, file processing exploits, such as those highlighted by Freiling et al.~\cite{6234406}, offer insights into similar architectural vulnerabilities.

Bypassing WAFs at the protocol level typically involves manipulating communication protocols. HTTP request smuggling, where a malicious request is embedded within a legitimate one, is a key example, with T-Reqs~\cite{treqs} providing methods to identify such vulnerabilities. Similarly, features such as chunked transfer encoding, which involves fragmenting HTTP data into smaller chunks, can be exploited to evade WAF detection. Dalili's work on IP fragmentation~\cite{soroush_dalili} explore these protocol-level evasion strategies. Also, SYMTCP~\cite{symtcp} focuses on evading deep packet inspection. Furthermore, Handley et al.~\cite{270920} inspects the evasion of network intrusion detection systems, a concept closely related to bypassing WAFs at the protocol level.

Recent work by Wang et al.~\cite{wang2024break} explores WAF bypasses through protocol-level evasion, whereas our study takes a distinct approach by targeting content-type-specific parsing discrepancies across headers and bodies. While both use fuzzing, our methodology systematically covers a wider range of content types, leading to the discovery of new bypass classes. We also propose a more detailed classification system with 24 attack types (versus 3 in prior work), validate our techniques on 100 real-world websites beyond controlled environments, and introduce a proof-of-concept normalizer to mitigate these issues. None of which are addressed in Wang et al.'s work.

Payload level bypassing techniques involve transforming the original payload to render it undetectable by the WAF. Obfuscation techniques, such as encoding or encryption, are frequently employed to conceal malicious code from WAF scrutiny. Additionally, evasion techniques, such as altering the order of characters or incorporating whitespace, can be used to confuse the WAF, and facilitate bypassing. Payload size manipulation, such as oversized POST requests that exploited a vulnerability in Google Cloud Armor~\cite{noauthor_google_2022}, illustrates how payload modifications can defeat WAF protections. Not all bypasses are feasible using by targeting a single level approach above. Some works involve combining multiple techniques to achieve a bypass. AutoSpear~\cite{qu_autospear_nodate} exemplifies a tool that automates WAF bypassing using a combination of architectural, protocol, and payload-level techniques. %

The academic works discussed here significantly enhance our knowledge of bypass techniques and WAF behavior, providing a platform for further exploration in WAF bypassing methods. Our work builds upon this foundation by introducing a novel approach that exploits content-parsing discrepancies, pushing the boundaries of existing methodologies. Unlike prior research, which often focuses on obfuscating payloads or manipulating protocol features, our approach targets fundamental weaknesses in how WAFs parse HTTP requests.

\section{\System Design and Methodology}

Figure~\ref{fig:diagram} provides an overview of the \System approach for finding discrepancy-based bypass vectors. The methodology involves generating, mutating, and testing HTTP requests across the WAFs and web application frameworks to identify successful bypasses.

\subsection{Input Generation and Mutation}
The fuzzer generates valid HTTP requests using a predefined grammar that contains a web attack payload (e.g., SQL injection, Cross-Site Scripting). Mutations are applied everywhere except the attack payload itself, ensuring that bypass attempts focus on discrepancies rather than payload obfuscation. The highlighted part in Listing~\ref{listing:normal-multi} shows an XSS attack payload, which remains unmodified during mutations. These mutations are expected to give us new discrepancy-based bypass instances. The goal is to confuse WAFs so they fail to properly parse the mutated requests and allow them through, while the web application frameworks, despite the mutations, correctly interpret and execute the attack payload.

\subsection{Identifying Successful Bypass Instances}

The target infrastructure consists of all tested WAFs where each WAF is able to forward requests to all web application framework instances. Each instance is a tiny web application that parses received requests by using the parser library of the web application framework (e.g., Flask) for the tested content type (e.g., \formdata). The results of parsing operation are logged for a later analysis. If a log shows that the request is parsed successfully and the request body contains the attack payload, then the corresponding input request can be used to successfully bypass the relevant WAF.

\subsection{Motivating Example}

Listing~\ref{listing:paramcon} presents an example of how parsing discrepancies can lead to WAF circumvention.
This example has two boundaries defined: \texttt{fake-boundary} and \texttt{real-boundary}. When this request arrives at the WAF, it picks the first one (i.e., \texttt{fake-boundary}) as the boundary value and ignores the remaining boundaries, thereby missing the attack payload. When it is forwarded to the web application framework running behind the WAF, as it supports the boundary continuation mechanism -- which was standardized in RFC 2231 and allows the use of multiple parameters to contain a single parameter value -- it takes the boundary value to be the concatenation of those two values (i.e., \texttt{real-boundary}), thereby parsing the XSS payload encapsulated between the \texttt{real-boundary} boundaries. 

\begin{lstlisting}[style=highlighted_http, caption={A malicious HTTP request using multipart/form-data with an XSS attack payload.}, label={listing:normal-multi}]
POST / HTTP/1.1
Host: victim.com
Content-Length: 114
Content-Type: multipart/form-data; (*\textcolor{blue}{boundary=1234}*)

--1234
Content-Disposition: form-data; name="field1"

(*\textcolor{red}{<script>alert(document.cookie)</script>}*)
--1234--
\end{lstlisting}

\begin{lstlisting}[style=highlighted_http, caption={A malicious \formdata request that uses parameter continuation.}, label={listing:paramcon}]
POST / HTTP/1.1
Host: victim.com
Content-Length: 230
Content-Type: multipart/form-data;
(*\textcolor{blue}{boundary=fake-boundary;boundary*0=real-;
boundary*1=}*)
(*\textcolor{blue}{boundary}*)

(*\textcolor{gray}{-\--fake-boundary}*)
Content-Disposition: form-data; name="field1"

value1
(*\textcolor{gray}{-\--fake-boundary-\--}*)
(*\textcolor{orange}{-\--real-boundary}*)
Content-Disposition: form-data: name="id"

(*\textcolor{red}{<script>alert(document.cookie)</script>}*)
(*\textcolor{orange}{-\--real-boundary-\--}*)
\end{lstlisting}

\section{Experiment Setup}

\subsection{Tools and Infrastructure for Experiments}
We used a modified version of the T-Reqs~\cite{treqs} fuzzer for the experiments. The modifications include adding support for encrypted HTTP requests, and setting \texttt{content-length} dynamically based on the length of the body. We also modified the T-Reqs source code to avoid using \texttt{<} and \texttt{>} characters to parse the grammar, since these characters were also used in the XSS attack payload that was present in generated requests. 

Input grammars were created in accordance with RFC specifications, encompassing all standard components of the content type and their possible values. The fuzzer was executed using these specific grammars\footnote{Available in the \href{https://github.com/sa-akhavani/waffled}{public repository}} for several hours, generating and testing a total of 373,670 requests across 5 WAFs and 6 web application frameworks.

Our server infrastructure was deployed across multiple cloud platforms, including Amazon AWS, Google Cloud, and Microsoft Azure. This configuration was utilized to host our web applications and conduct the experiments.

\subsection{WAF Configuration and Setup}
To comprehensively assess the resilience of WAFs against mutated HTTP/1.1 web requests, we conducted an extensive evaluation of several WAFs. The selection criteria focused on solutions with significant adoption across both commercial and open-source domains. The WAFs included in this study were AWS WAF, Cloudflare WAF, Google Cloud Armor, Microsoft Azure WAF, and ModSecurity on NGINX. This selection enabled us to evaluate the effectiveness of both widely-used community-driven tools and leading commercial solutions in the industry.

For all tested WAFs, default settings were utilized without any modifications to preserve the integrity of the testing environment and ensure the results accurately reflected typical user experiences. The consistent configuration, especially with the widespread adoption of the OWASP CRS, enabled a fair comparison across WAFs despite their differing underlying technologies and implementation strategies. This standardized methodology established a clear baseline to evaluate each WAF's capability to process mutated HTTP requests, effectively highlighting their individual strengths and potential vulnerabilities.

\begin{table}[t]
    \centering
    \caption{List of Tested WAFs and Their Supported Request Content-Types (As of July 2024).}
    \label{tab:wafs}
    \begin{tabular}{llc}
        \toprule
        \textbf{WAF} & \textbf{Content-Type} & \textbf{Bypassed} \\ \midrule
        Cloudflare WAF & multipart, json, xml & Yes \\ 
        Microsoft Azure WAF & multipart, json & Yes \\ 
        Google Cloud Armor & multipart, json & Yes \\ 
        Amazon AWS WAF & multipart, json & No \\ 
        ModSecurity on NGINX & multipart, json, xml & Yes \\ \bottomrule
    \end{tabular}
\end{table}

Table~\ref{tab:wafs} lists the WAFs and their supported content-types in our study.

\noindent \textbf{AWS WAF.} AWS WAF was configured using Elastic Load Balancing (ELB). We employed the AWS Managed Ruleset in conjunction with OWASP CRS version 3.0, using default settings to accurately represent typical user configurations.

\noindent \textbf{Cloudflare WAF.} Cloudflare WAF was set up as the DNS provider for our domain, applying its WAF to all incoming traffic via its global CDN. We used the Pro version, enabling both Cloudflare's Managed Ruleset and the OWASP CRS to ensure comprehensive threat coverage.

\noindent \textbf{Google Cloud Armor.} Google Cloud Armor was deployed on a virtual machine behind Google Cloud's Load Balancer. We configured it with rulesets for SQL injection and XSS protection (\texttt{sqli-v33-stable} and \texttt{xss-v33-stable}) at Sensitivity Level 1.

\noindent \textbf{Microsoft Azure WAF.} Microsoft Azure WAF was set up through Azure's Application Gateway using the WAF V2 tier, with OWASP CRS version 3.0 enabled by default. This configuration aligns with common deployment practices.

\noindent \textbf{ModSecurity.} ModSecurity was deployed as a plugin for NGINX, configured with OWASP CRS version 3.0 at Paranoia Level 1, reflecting the default settings commonly used in many environments. The setup adhered to the official documentation guidelines, ensuring an accurate evaluation of its effectiveness in scenarios typical of real-world deployments.

The influence of rulesets on our findings is minimal for several key reasons. First, the attack payloads used in this study are intentionally straightforward, designed to be blocked by all standard rulesets across the tested WAFs, as confirmed during preliminary testing. Second, our bypass techniques do not depend on obfuscating the payload itself, but rather exploit parsing inconsistencies, focusing on how WAFs interpret HTTP requests before applying their rulesets. This approach highlights that the bypasses stem primarily from the WAF's inability to correctly parse the request or its decision to bypass parsing due to the complexity of the input, irrespective of the specific ruleset in use.

\subsection{Web Application Frameworks Setup}

In this work, we evaluated seven widely-used web application frameworks: Express, Node.js-HTTP, Flask, FastAPI, Gin, Laravel, and Spring Boot. To comprehensively analyze their parsing capabilities, we employed a combination of the frameworks' default parsers and HTTP parser packages sourced from reputable platforms. The selection of parser packages was informed by their prevalence and popularity within the developer community, as indicated by metrics from resources such as npm and PyPi Stats. This approach ensured that our study included modules that are both widely adopted and representative of real-world usage.

To accurately test how each framework and parser handles the request body, we implemented the following approach: we attempted to access the parsed content of the request body through framework-specific methods such as \texttt{request.body.parse()}, etc. The goal was to retrieve the value of a specific field (\texttt{field}), and have the framework and parser process the entire request body. After parsing, we searched for the sent attack payload within the parsed content. If any of the parsed fields contained the attack payload, it indicated that the request successfully bypassed the WAF, and was correctly parsed by the framework and parser. We then sent a success flag along with instance name and WAF information back to our client to mark the successful attacks.

For parsing request bodies in our study, each framework's default parser was utilized. In cases where the default parser did not support a specific content-type, we employed the most popular third-party parsers for that content-type. This approach of using both default and third-party parsers provided an understanding of how web application frameworks handle and interpret HTTP requests under different parsing scenarios.

For Node.js, the chosen parsers included \texttt{Busboy} and \texttt{Formidable} for parsing multipart content types, and \texttt{fast-xml-parser} for handling \texttt{application/xml} requests. Since Express and the HTTP module do not support parsing of multipart and xml content-types.
For Python frameworks, we integrated \texttt{xmltodict} and \texttt{xmlminidom} for XML parsing. Multipart content types in Python were processed using \texttt{python-multipart} parsers.

In all our tests, we strictly adhered to the official documentation and examples of each framework, avoiding any custom development. This ensured that our evaluation covered the most common and correct usage patterns of these frameworks and parsers.

\subsection{Validation of Bypass Effectiveness}
To validate WAF bypasses, we dispatched malicious requests through the WAF to target frameworks, which attempted to parse and store the body content on the server. We then compared the original payload with the stored result to determine if the bypass succeeded. A script automated this comparison, confirming whether the payload had both circumvented the WAF and been correctly parsed by the framework. This process was necessary because defining a successful bypass solely based on the ability of a request to evade the WAF would lead to misclassification. 
Such a definition could include malformed requests that are not conforming to HTTP standards or are unprocessable by any framework, as successful bypasses. This would lead to inflating false positives and obscuring real-world risks because of the following reasons:

\textbf{Unrealistic Attack Scenarios.} Most WAFs allow malformed requests to pass because they cannot parse them. If we were to consider these instances as bypasses, more than 50\% of mutated requests could be misclassified as successful.

\textbf{Practical Mitigation Challenges.} Malformed requests are frequently rejected at the framework level. Reporting these would inflate false positives and obscure actionable findings.

A true attack occurs when a malicious request bypasses the WAF, is parsed by the framework, and the payload is executed. Malformed requests that frameworks cannot parse pose lower security risks and are lower priority. All bypasses in our work meet this criteria, enabling execution of the malicious payload (e.g., XSS or SQL injection).

\subsection{Experiment Results and Analysis}
Table~\ref{tab:frameworks} provides the list of frameworks and Table~\ref{tab:parsers} provides the list of parsers that were inspected in our work. The experiments revealed that all of the frameworks and parsers were vulnerable to at least one bypass technique, highlighting the effectiveness and practical implications of our findings.

\begin{table}[t]
    \centering
    \caption{List of tested web frameworks and whether they are vulnerable to at least one bypass in our study or not.}
    \label{tab:frameworks}
    \begin{tabular}{ccccc}
        \toprule
        \textbf{Framework} & \textbf{Version} & \textbf{Language} & \textbf{Vulnerable} \\ 
        \midrule
        Laravel & 10.48.16 & PHP & Yes \\ %
        Spring Boot & 3.2.2 & Java & Yes \\ %
        Gin & v1.9.1 & Go & Yes \\ %
        Express & 4.18.2 & Node.js & Yes \\ %
        Fastapi & 0.109.2 & Python & Yes \\ %
        Flask & 3.0.2 & Python & Yes \\ %
        Node.js-HTTP & 18.16.1 & Node.js & Yes \\ 
        \bottomrule
    \end{tabular}
\end{table}

\begin{table}[ht]
    \centering
    \caption{List of tested parsers, their weekly downloads as of July 2024 collected from NPM and PyPI Stats~\protect\footnotemark website, and whether they are vulnerable to at least one bypass in our study or not.}
    \label{tab:parsers}
    \begin{tabular}{ccccc}
        \toprule
        \textbf{Parser} & \textbf{Version} & \textbf{Vulnerable} & \textbf{Downloads} \\ \midrule
        \multicolumn{1}{c}{\textbf{Node.js}} & \\ \cmidrule(l){1-1} %
        Busboy & 1.6.0 & Yes & 8,639,607 \\ %
        Formidable & 3.5.1 & Yes & 6,998,735 \\ %
        fast-xml-parser & 4.3.3 & Yes & 14,254,527\\ \midrule
        \multicolumn{1}{c}{\textbf{Python}} & \\ \cmidrule(l){1-1}
        xmltodict & 0.13.0 & Yes & 10,374,671 \\ %
        xmlminidom & 3.10.12 & Yes & n/a \\ %
        python-multipart & 0.0.9 & Yes & 4,288,034 \\ \bottomrule
    \end{tabular}
\end{table}

\footnotetext{\url{https://www.npmjs.com} and \url{https://pypistats.org}}

\section{Findings}

Before diving into analyzing these bypasses and reporting them, we first, analyze all results to retain only those bypassed requests that are achieved with the minimum number of mutations. Then, we classify bypasses based on their mutations. This allows us to:
\begin{itemize}
    \item Identify the minimum mutations needed for a request to bypass a WAF uniquely.
    \item Ensure our results are realistic, and our reports are minimal and accurate by focusing solely on unique bypasses.
    \item Reporting and documenting bypass techniques and ensuring that mitigation strategies can address entire classes of bypasses rather than individual instances.
\end{itemize}

\subsection{Identification of Unique Bypass Requests}
In the process of testing and validating bypass techniques against WAFs, identifying unique bypass requests is crucial for accurately assessing the effectiveness of the discovered vulnerabilities.
The identification of unique bypass requests involves rigorous analysis and filtering of the mutated HTTP requests generated during the testing phase.

Not all successfully bypassed requests are unique in terms of the underlying techniques used. For instance, consider a scenario where two mutations contribute to the bypass success:

\begin{itemize}
    \item \textbf{Mutation I}: Introduces a \texttt{\textbackslash x00} character into the boundary value of a multipart request, transforming \texttt{--boundary} to \texttt{--\textbackslash x00boundary}.
    \item \textbf{Mutation II}: Changes the capitalization of characters in the \texttt{charset} parameter of the \texttt{Content-Type} header from \texttt{utf-8} to \texttt{Utf-8}.
\end{itemize}

While Mutation I, alone, constitutes a successful bypass strategy, Mutation II does not contribute to bypassing the WAF independently. However, when both mutations are combined into a single request, the request successfully evades WAF detection due to the presence of Mutation A. In this scenario, the unique bypass is attributed to Mutation A, whereas Mutation B does not qualify as an independent bypass strategy. And the request that is formed by both of these mutations is not a unique bypass.

To maintain the integrity and accuracy of the findings, redundant bypass requests that do not introduce unique evasion techniques are filtered out during the analysis phase. 

After applying this minimization technique, among all of the generated requests, 1207 of them were unique bypasses found for (WAF, Framework) pairs.

\subsection{Classification of Bypasses}

After extracting the bypasses with the minimum number of mutations, we proceed with the classification of these bypasses. This classification process is essential as it allows us to systematically analyze and categorize the bypasses, facilitating the development of effective defense mechanisms.

To achieve this, each bypass is examined based on the unique mutation strategy and the mutated element used rather than the specific mutation. For instance, consider a JSON request with the body \texttt{\{"field1": "value1"\}}. If two distinct bypasses are achieved by inserting a \texttt{\textbackslash x00} or \texttt{\textbackslash x02} after the second double quote, both are classified under the same category: \textit{manipulating the field name wrapper}. Although different characters are used, the fundamental concept remains consistent. This categorization simplifies the study of bypass techniques by grouping similar strategies, thereby enabling a more structured approach to understanding and mitigating these vulnerabilities.

\subsection{Bypassed Requests Analysis}
We now present the classification results of the bypassed requests for each content-type in our work. Some of the mutation classes are common among all tested content-types, they are defined once but mentioned for each content-type that they were leading to a successful bypass.

\begin{table*}[h]
    \centering
    \caption{Classification of Multipart Bypass Categories with Examples. Removals are displayed with a strike-through text with a gray background, while additions and replacements are shown with only a gray background.}
    \begin{tabular}{cc}
        \toprule
        \textbf{Category Name} & \textbf{Request Example} \\
        \midrule
        Boundary Delimiter Manipulation & \texttt{\colorbox{codegray}{\textcolor{red}{\st{\textbackslash r\textbackslash n}}}---boundary} \\
        \hline
        Content-Disposition Disruption & \texttt{content-disposition: form-da\colorbox{codegray}{\textcolor{blue}{\textbackslash x00}}a;} \\
        \hline     
        Distorted Header Injection to Body & \texttt{\colorbox{codegray}{\textcolor{blue}{conten\textbackslash x00-extra: something}}} \\
        \hline        
        Content-Type Tweak in Body & \texttt{Content-Type: text/plain\colorbox{codegray}{\textcolor{blue}{\textbackslash x00}}; charset=UTF-8} \\
        \hline
        Charset Value Alteration in Body & \texttt{charset=\colorbox{codegray}{\textcolor{blue}{\textbackslash x00}}UTF-8} \\
        \hline
        Header Separator Manipulation in Body & \texttt{content-disposition: form-data; name="f1"\colorbox{codegray}{\textcolor{blue}{\textbackslash x00}}} \\
        \hline        
        Content-Type Parameter Tweak & \texttt{C\colorbox{codegray}{\textcolor{red}{\st{o}}}ntent-Type: multipart/form-data;} \\        
        \hline
        Boundary Delimiter Removal & \texttt{\colorbox{codegray}{\textcolor{red}{\st{---boundary}}}} \\
        \hline      
        Linefeed Removal & \texttt{Content-Type: multipart/form-data; boundary=real\textbackslash r\colorbox{codegray}{\textcolor{red}{\st{\textbackslash n}}}\textbackslash r\textbackslash n} \\
        \hline

        Whitespace Alteration & \texttt{
        Content-Type: \colorbox{codegray}{\textcolor{blue}{\textbackslash t}}multipart/form-data; boundary*0=re;boundary*1=al} \\
        \hline

        Disrupted Body Field & \texttt{
        content-disposition: form-data; name="field1\colorbox{codegray}{\textcolor{blue}{\textbackslash x00}}"} \\
        \hline        
        Boundary Header Tampering & \texttt{---boundary=value\colorbox{codegray}{\textcolor{blue}{;}}} \\

        \bottomrule

    \end{tabular}
    \label{tab:multipart_bypass_categories}
\end{table*}

\subsubsection{Multipart Bypass Classes}
\label{subsec:multipart_classes}
\begin{figure}[t]
    \centering
    \includegraphics[width=1\columnwidth]{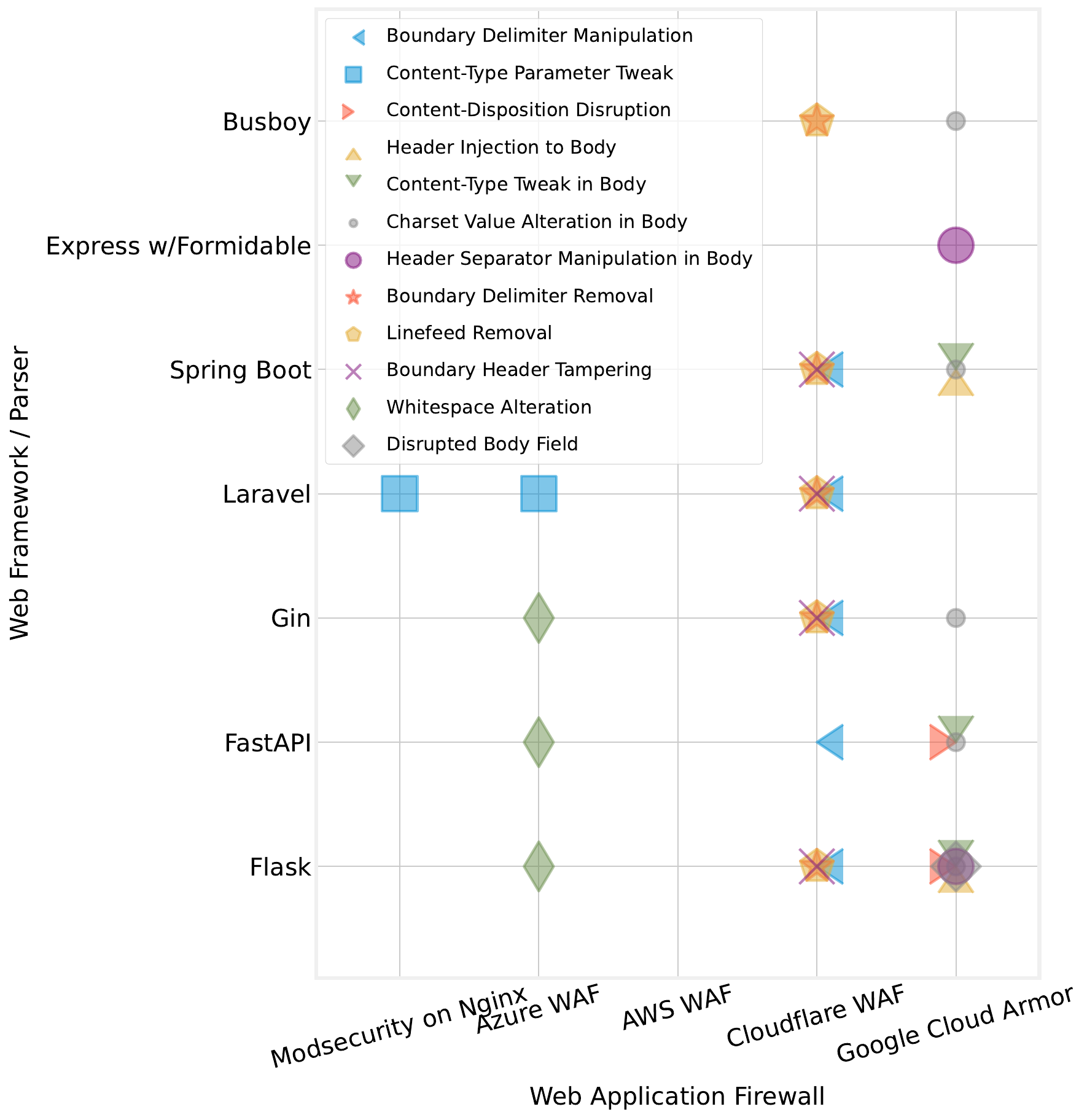}
    \caption{Occurrence of Successful Bypasses For Multipart Content-Types in Tested Frameworks}
    \label{fig:multipart_bypass_distribution}
    \vspace{-2.0ex}
\end{figure}

Our examination identified a total of 351 unique bypasses related to multipart content-type parsing. Figure~\ref{fig:multipart_bypass_distribution} demonstrates valid bypass classes that we found in our work for multipart content-type. Table~\ref{tab:multipart_bypass_categories} contains examples for each discussed bypass category.

\textbf{Boundary Delimiter Manipulation.} This category involves the removal or alteration of boundary delimiters within multipart content. For instance, removing the \texttt{\textbackslash r\textbackslash n} sequence before the boundary string can confuse the parsing logic, leading to successful bypasses.

\textbf{Content-Type Parameter Tweak.} This technique includes modifications to the global Content-Type header's name, such as removing or inserting characters.

\textbf{Content-Disposition Disruption.} This technique targets the Content-Disposition header within the multipart content, altering its structure to evade detection.

\textbf{Disrupted Header Injection to Body.}
This category includes adding redundant headers with disrupted header name directly into the multipart body content, which can mislead the WAF into processing the content incorrectly.

\textbf{Content-Type Tweak in Body.}
Here, manipulations are performed on the Content-Type value within the body and not in the global header, such as inserting characters.

\textbf{Charset Value Alteration in Body.}
This involves altering the charset value within the body content, affecting how the WAF interprets the encoding of the payload.

\textbf{Header Separator Manipulation in Body.} This technique modifies the separator between multiple header lines in the multipart body content, such as replacing newlines with other characters.
    
\textbf{Boundary Delimiter Removal.}
This category involves the complete removal of boundary delimiters, affecting the WAF's ability to correctly parse multipart sections.

\textbf{Linefeed Removal.}
This technique removes newline and carriage return characters. For example, removing the linefeed after the request headers and before beginning the boundary delimiter for the multipart request body. This approach can cause the WAF to misinterpret the structure of the multipart content.

\textbf{Boundary Header Tampering.} In this category, manipulations involve changes to the boundary value in the header, such as appending a semicolon to the boundary parameter in the content-type header line.

\textbf{Whitespace Alteration.} In this category, the whitespace character is replaced with \texttt{\textbackslash t} in the content-type header.
The request header must use the parameter value continuation feature of HTTP to define the boundary for this bypass to be effective.
\texttt{boundary*0=re;boundary*1=al}

\textbf{Disrupted Body Field.} In this category, manipulations involve inserting invalid characters in a field name. The request header must use the parameter value continuation feature of HTTP to define the boundary for this bypass to be effective.

\subsubsection{XML Bypass Classes}
\begin{figure}[t]
    \centering
    \includegraphics[width=\columnwidth]{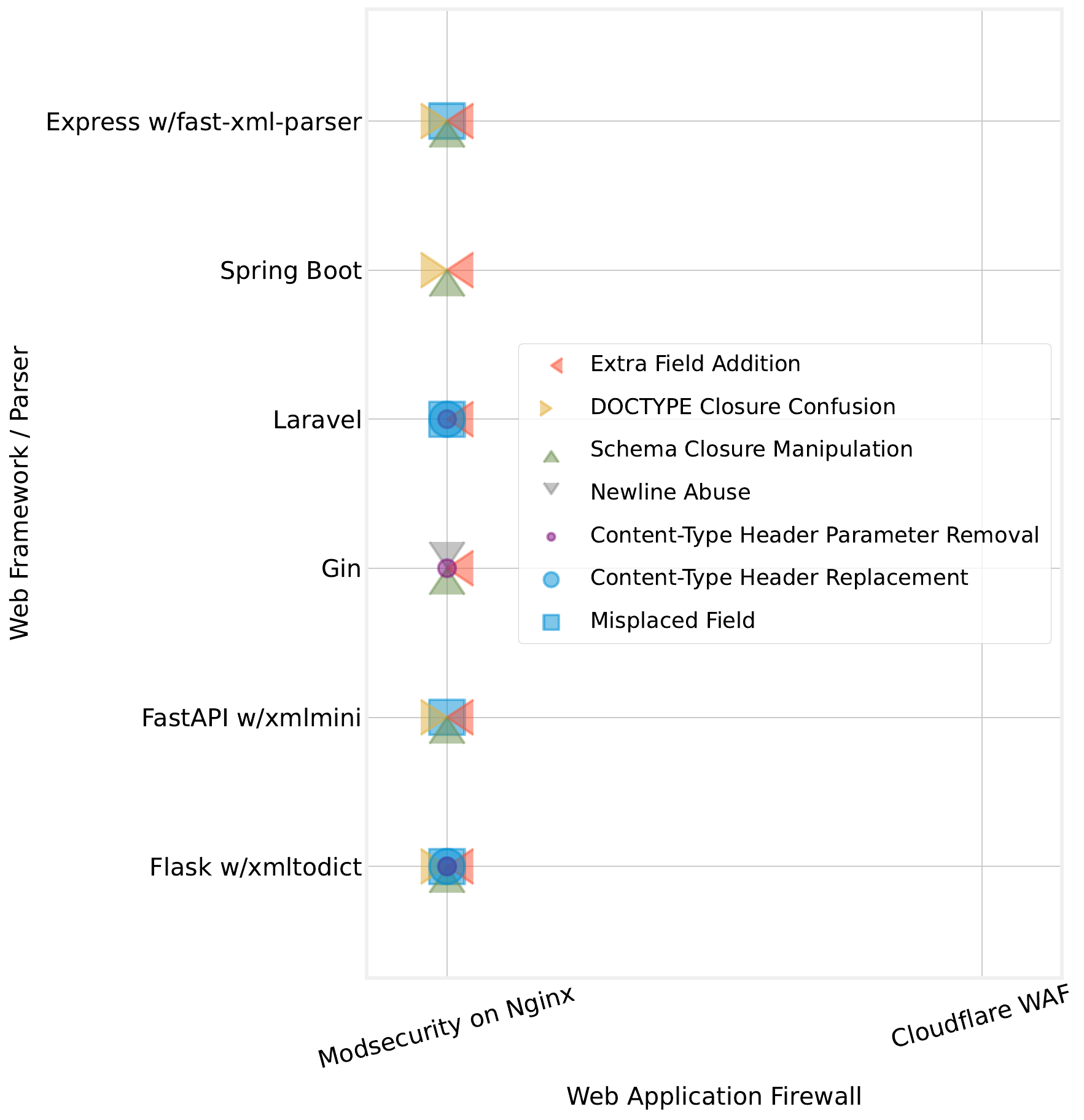}
    \caption{Occurrence of Successful Bypasses For application/xml Content-Types in Tested Frameworks}
    \label{fig:xml_bypass_distribution}
    \vspace{-2.0ex}
\end{figure}

Our analysis revealed a total of 299 unique bypasses for XML content types. Figure~\ref{fig:xml_bypass_distribution} demonstrates valid bypass classes for application/xml content-type. Table~\ref{tab:xml_bypass_categories} contains examples for each discussed bypass category.

\textbf{Extra Field Addition.} Adding an extra field outside the defined XML schema. For example, inserting a new field such as \texttt{<field2>value2</field2>} exploits schema validation weaknesses, allowing malicious payloads to bypass security checks.
    
\textbf{DOCTYPE Closure Confusion.} Placing an extra character at the end of the xml body confuses the WAF in parsing the DOCTYPE entity of the XML.

\textbf{Schema Closure Manipulation.} This technique manipulates the closure of XML schemas. By inserting characters, new elements, or duplicated field names and values at specific positions within the schema, the structure of the XML document is altered in a way that evades detection.
    
\textbf{Newline Abuse.} This method achieves the bypass by placing an extra new-line before the content-type header.

\textbf{Content-Type Header Parameter Removal.} Bypasses achieved by removing the Content-Type header's name. Such manipulations exploit the dependency of some WAFs on specific header configurations to enforce security rules.

\textbf{Content-Type Header Replacement.} The parameter name within the Content-Type header is replaced with its value. This manipulation can confuse WAFs that rely on precise header structures for detection.

\begin{figure}[ht]
    \centering
    \includegraphics[width=\columnwidth]{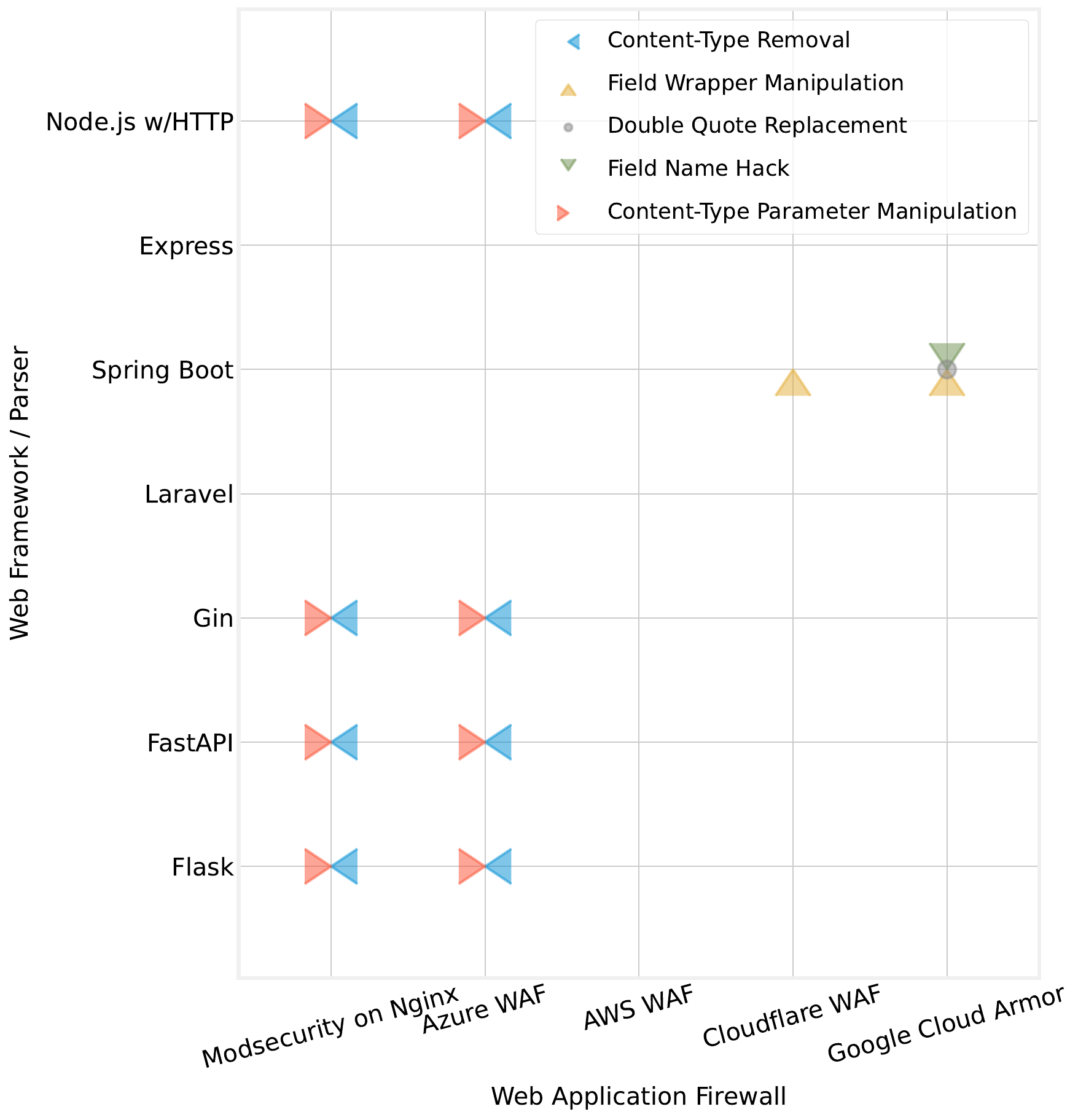}
    \caption{Occurrence of Successful Bypasses For application/json Content-Types in Tested Frameworks}
    \label{fig:json_bypass_distribution}
    \vspace{-2.0ex}
\end{figure}

\textbf{Misplaced Field.} Placing field values outside their corresponding XML tags. By altering the position of values within the XML document, the attack payload can bypass WAF rules that expect a specific structure.

\begin{table*}[h]
    \centering
    \caption{Classification of XML Bypass Categories with Examples}
    \begin{tabular}{cc}
        \toprule
        \textbf{Category Name} & \textbf{Request Example} \\
        \midrule
        Extra Field Addition & \texttt{<field1>value1</field1>\colorbox{codegray}{\textcolor{blue}{<field2 attr="history">hi</field2>}}} \\
        \hline
        DOCTYPE Closure Confusion & \texttt{<!DOCTYPE BOOK [...]><field1>value1</field1>\colorbox{codegray}{\textcolor{blue}{]}}</BOOK>} \\
        \hline
        Schema Closure Manipulation & \texttt{<genre:schema><field1>value1</field1>\colorbox{codegray}{\textcolor{blue}{j}}</genre:schema>} \\
        \hline
        Newline Abuse & \texttt{\colorbox{codegray}{\textcolor{blue}{\textbackslash r\textbackslash n}}Content-Type: application/xml} \\
        \hline
        Content-Type Header Parameter Removal & \texttt{\colorbox{codegray}{\textcolor{red}{\st{Content-Type: }}}application/xml} \\
        \hline
        Content-Type Header Replacement & \texttt{\textcolor{blue}{\colorbox{codegray}{application/xml}}application/xml} \\
        \hline
        Misplaced Field & \texttt{...\colorbox{codegray}{\textcolor{blue}{value1}}<field1>value1</field1>...} \\
        \bottomrule
    \end{tabular}
    \label{tab:xml_bypass_categories}
\end{table*}

\subsubsection{JSON Bypass Classes}
In our work, we identified a total of 557 unique bypasses for JSON content types. Figure~\ref{fig:json_bypass_distribution} demonstrates valid bypass classes for application/json content-type. Table~\ref{tab:json_bypass_categories} contains examples for each discussed bypass category.

\textbf{Content-Type Removal.} This category involves bypassing the WAF by removing the \texttt{Content-Type: application/json} header from the global header. This manipulation exploits the reliance of some WAFs on specific content-type headers to enforce security rules.

\textbf{Field Wrapper Manipulation.} In this category, the attack is achieved by adding characters such as \texttt{\textbackslash x00} between the field name and the colon in the JSON object. This manipulation disrupts the field parsing.

\textbf{Double Quote Replacement.} This method replaces the double quotes surrounding the field names and values with  other characters.

\textbf{Field Name Hack.}
This technique involves altering characters within the field names of the JSON object. For example, replacing a character in the field name with \texttt{"\textbackslash x00"} can bypass WAFs that do not properly handle such anomalies in field names.

\textbf{Content-Type Parameter Manipulation.} Here, the manipulation targets the Content-Type header name.

\begin{table*}[h]
    \centering
    \caption{Classification of JSON Bypass Categories with Examples.}
    \begin{tabular}{cc}
        \toprule
        \textbf{Category Name} & \textbf{Bypassed Request Example} \\
        \midrule
        Content-Type Removal & \texttt{\colorbox{codegray}{\textcolor{red}{\st{Content-Type: application/json}}}} \\
        \hline
        Field Wrapper Manipulation & \texttt{\{ "field1" \colorbox{codegray}{\textcolor{blue}{\textbackslash x00}}: "<script>alert(document.cookie)</script>" \}} \\
        \hline
        Double Quote Replacement & \texttt{\{ "field1\colorbox{codegray}{\textcolor{blue}{\textbackslash x00}}: "<script>alert(document.cookie)</script>" \}} \\
        \hline
        Field Name Hack & \texttt{\{ "f\colorbox{codegray}{\textcolor{blue}{\textbackslash x00}}eld1": "<script>alert(document.cookie)</script>" \}} \\
        \hline
        Content-Type Parameter Manipulation & \texttt{Content\colorbox{codegray}{\textcolor{red}{-}}Type: application/json} \\
        \bottomrule
    \end{tabular}
    \label{tab:json_bypass_categories}
\end{table*}

\subsection{Practicality of Findings}
The bypass findings of this paper for three content types are practical and useful only if these content types are popular amongst the web applications in the real world. We performed a study on a sample collected from real-world applications to gauge the popularity of these content types. In addition, we performed another study to check interchangeability between the content-types which can also contribute to the practicality of our findings.

We collected 100 high-ranking websites from PublicWWW~\footnote{\url{https://publicwww.com/}}, which is a search engine for website source codes, by using the query below which was tailored for the needs of the study.
\newline
\mbox{\texttt{type="email" "forgot password" -reCAPTCHA}}

This search query allowed us to find ``forgot-password'' web pages which contain an HTML form with an email-type input field and does not contain a CAPTCHA. The presence of CAPTCHA does not allow replaying the ``forgot-password'' requests, which makes it easier and faster to check the interchangeability of content types. We chose the first 100 search results, which are the ``forgot-password'' pages of popular websites ranking between 3K-50K. About half of these webpages were excluded from the study for reasons such as ``non-english language'', ``obscenity'' and ``an emerging captcha''.

On each ``forgot-password'' page, we submitted the form after typing in a non-existent email address and intercepted the generated HTTP request on the Burp Suite tool. We examined the \texttt{Content-Type} header field and the request body format to decide the content-type for each request. We found that more than two-thirds of the websites use the \texttt{application/x-www-form-urlencoded}, while one-fourth of them use the \texttt{application/json} and its variation. Only two websites use \texttt{multipart/form-data} and one uses URL parameters. For each intercepted ``forgot-password'' request, to check the interchangeability between content types, we used the ``Change body encoding'' feature of Burp Suite (in the ``Repeater'' section) to convert the content type to \texttt{multipart/form-data} in \texttt{application/x-www-form-urlencoded} requests. We then examined the response status code and response body to check the sameness of responses, which indicates that the application does not differentiate those two content types. We found that more than ninety percent of websites accept them both.

These results suggest that our bypass findings are widely applicable and practical across web applications. The largest portion (i.e., more than two-thirds), which uses \texttt{application/x-www-form-urlencoded}, are affected by our \texttt{multipart/form-data} findings, since an attacker can easily switch the content type and apply the bypass techniques. The second largest portion, which uses \texttt{application/json}, is affected by all the relevant bypass findings we reveal in this paper. The fact that these two content types together make up more than 90\% of content types used across web applications, clearly shows the extent of the practicality and impact of these bypass techniques.

\section{Protecting WAFs and Frameworks by Normalizing HTTP Request Bodies}

In this section, we discuss our approach to mitigate the identified vulnerabilities by introducing an HTTP multipart body normalizer that protects web application firewalls (WAFs) and web application frameworks from parsing-related attacks. The normalizer acts as a gateway for the WAF that strictly enforces RFC grammar rules related to multipart form data, and rejects requests that are not compliant. For those requests that are compliant, optional message body components and fields that must be ignored are removed. This ensures that problematic message constructs are removed before they can induce differential parsing behavior between the WAF and the web application framework. The \normalizer's functionality could be expanded for each content-type to provide protection against other popular content types as well. We are introducing a proof of concept description in this section, and use \formdata as demonstration.

\noindent \textbf{Overview of HTTP Normalizer.}
The HTTP multipart body normalizer serves two primary functions: normalizing multipart request bodies, and rejecting non-compliant requests.  Upon receiving an HTTP request, the normalizer separates the request's headers and body using the AIOHTTP Python library. It then parses and validates the request's \texttt{Content-Type} header, and if it indicates that the message body uses the MIME multipart encoding, it performs one of the following actions:

\begin{itemize}
    \item \textbf{Normalization.} The normalizer begins by parsing requests using a strict multipart MIME parser generated from the ABNF in the RFCs. The output data structure of the parser is capable of representing only the necessary components of a multipart message body. Deprecated and optional portions of the multipart message are not representable in this structure. The structure is then reserialized and forwarded to its destination. Crucially, the normalized request is deserialized from a data structure in which invalid state is not representable, so the output of the normalizer is never malformed.
    \item \textbf{Rejection.} If the normalization process cannot be completed because necessary components of the request body uses a deprecated RFC feature or is malformed, the request is rejected.
\end{itemize}

Listing~\ref{listing:no_normal} demonstrates a sample malformed request before being passed to the normalizer. Listing~\ref{listing:normalized} demonstrates how the sample request would look like after being normalized.

\begin{lstlisting}[style=highlighted_http, caption={A request before normalization.}, label={listing:no_normal}, float]
POST / HTTP/1.1
Host: target.com
Content-Type: multipart/FoRm-dAtA; boundary="1234"
Content-Length: 90

--1234
Content-DISPOSITION:\tform-data;name="files";\t filename="ab.txt"

Foo
--1234--
\end{lstlisting}

\begin{lstlisting}[style=highlighted_http, caption={The same request after normalization. Note the standardized capitalization and spacing, removed trailing line ending, and inserted Content-Type MIME header.}, label={listing:normalized}, float]
POST / HTTP/1.1
Host: target.com
Content-Type: multipart/form-data; boundary=1234
Content-Length: 90
Accept: */*
Accept-Encoding: gzip, deflate
User-Agent: Python/3.12 aiohttp/3.9.5

--1234
Content-Type: text/plain
Content-Disposition: form-data; name="files"; filename="ab.txt"

Foo
--1234--
\end{lstlisting}

\noindent \textbf{Normalizing Request Body.}
Currently, the normalizer applies only to requests that use a multipart transfer encoding. Future enhancements will extend this normalization capability to other content-types, aiming to demonstrate the viability of this approach in enhancing WAF security by rejecting malicious HTTP requests and reducing false positives through normalization. This methodology can be uniformly applied to all content types to achieve comprehensive normalization.

\noindent \textbf{Evaluation of Normalizer.}
We evaluated the normalizer first by verifying that it does not reject any of a set of valid multipart message bodies. We obtained these from the test suites of the Tornado web server, RStudio, and the multiparty multipart message parser.

We sampled bypasses from all 12 discovered bypass classes that we mentioned in subsection~\ref{subsec:multipart_classes} and took 63 bypassable requests in total to test against the normalizer, and recorded whether the request was accepted, normalized, or rejected. Note that while we could mutate HTTP requests using our fuzzer and send all of them to the normalizer, it was unnecessary because: 1) even if a request bypassed the normalizer, it would have been blocked by the WAF since it was not among the found bypasses, and 2) the normalizer will ultimately be merged into existing WAF rulesets and does not need to be fuzzed separately because it is not a WAF itself, but a ruleset enforcer. 

Table~\ref{tab:evaluation} demonstrates these evaluation metrics. Out of the 8 bypassable requests that are not rejected by the normalizer, all are blocked by Cloudflare's WAF, resulting in a 100\% success rate.

\begin{table}[t]
    \centering
    \caption{Evaluation Metrics for HTTP Normalizer.}
    \begin{tabular}{ccc}
    \toprule
    & \textbf{Bypassable Requests} \\ 
    \bottomrule
    Normalized & 8 \\ %
    Rejected & 55 \\ 
    \hline
    Total Attempts & 63 \\ %
    \bottomrule
    \end{tabular}
    \label{tab:evaluation}
\end{table}

\noindent \textbf{Performance Cost.}
The \normalizer was developed to demonstrate that all identified bypass techniques can be prevented if WAFs adhere to proper parsing standards. Our findings, including the lack of successful bypasses against AWS WAF, validate this hypothesis. The intent of the project was not to create a fully optimized, production-grade tool with minimal overhead and broad content-type support, but rather to prove that an effective solution is feasible and implementable for a specific content type. By strictly adhering to RFC grammar rules, the HTTP-Normalizer effectively eliminates all bypass methods. While it currently functions as an additional evaluation layer, its rules can be seamlessly integrated into existing WAF rulesets with minimal performance impact.

\section{Discussion}
\noindent \textbf{Usability vs. Security Debate.}
While WAFs can prevent these bypasses by following RFC standards, real-world deployments may face compatibility or customer-specific constraints. Thus, we do not blame vendors, as they often balance security with operational needs. However, our proposed defense mechanism, the \normalizer, shows that all bypasses in this work are theoretically preventable without significantly impacting usability. Also, not all WAFs are vulnerable to every bypass in this research, indicating that a well-designed RFC-compliant parser can effectively mitigate such issues.

\noindent \textbf{Practical Considerations.}
We focused on prominent web frameworks and their default or most popular parser packages. While multipart forms are common, not all frameworks have dedicated parsers for them, often leading to custom implementations from developers. These custom parsers can increase the risk of developer errors, potentially allowing bypassed requests to be parsed successfully server-side. However, our work did not examine custom parsers, leaving their implications outside the scope of our findings.

\noindent \textbf{Limitations and Future Work.}
This work primarily focuses on HTTP/1.1 web requests, and future research could investigate bypass behaviors in HTTP/2. Additionally, while we covered major frameworks and WAFs, expanding the range of frameworks and content-types tested could provide broader insights. Our findings also highlight that parsing discrepancies can reveal fingerprinting opportunities for both WAFs and frameworks. Prior research has shown that HTTP parsing behavior can be used to fingerprint web servers~\cite{untangle2024}, and our results show similar potential for WAFs. For example, sending a JSON bypass request under the Field Name Manipulation category can help identify Google Cloud Armor with the Spring Boot framework, as only that combination would parse and accept the payload. Future work could develop a systematic fingerprinting methodology based on these discrepancies.

\section{Ethics and Disclosure Information}

All attacks and bypasses in this paper were conducted in a controlled environment using our infrastructure, ensuring no external impact. No malicious payloads were sent to the tested websites, and each site received no more than 10 requests. All WAF tests adhered to bug bounty protocols, and bypasses were disclosed to the affected vendors, who subsequently acknowledged the existence of these vulnerabilities.

\begin{itemize}
    \item \textbf{Google Cloud Armor} classified our report as a Tier 1, Priority 1, Severity 2 vulnerability under the "Insecure by Default" category and rewarded us with a bug bounty.
    \item \textbf{ModSecurity} acknowledged all bypasses we reported in CRS 3.3.
    \item \textbf{Cloudflare} confirmed the reported bypasses and stated that they are working on a fix.
    \item \textbf{Microsoft Azure} acknowledged the bypasses in CRS 3.0, which was the default ruleset at the time of our study. However, they are retiring CRS 3.0 and transitioning to DRS 2.1, an enhanced ruleset based on CRS 3.2, which addresses these vulnerabilities.
\end{itemize}

\section{Conclusion}
This work has highlighted the significant impact of parsing discrepancies on the effectiveness of Web Application Firewalls (WAFs) in protecting against cyber threats. Our experiments across major WAFs, particularly with \formdata, \json, and \xml content-types, resulted in 1207 bypasses. These bypasses were successfully parsed by our target web application frameworks employing default or popular request parsers. This finding indicates a critical vulnerability in WAFs' ability to uniformly interpret and filter web requests.

This paper's investigation into popular WAFs, including Cloudflare, Cloud Armor, AWS WAF, Azure WAF, and ModSecurity, highlights significant concerns regarding parsing discrepancies in their request analysis. These discrepancies pose a substantial risk, as WAF users may falsely believe they are protected against common attack payloads. As attackers employ increasingly sophisticated methods, the development of dynamic, intelligent WAFs becomes important. Our research not only contributes to the understanding of current WAF limitations, but also proposes effective solutions to defend against the identified threats. This work emphasizes the critical need for improved parsing consistency in WAFs to ensure robust protection against these attacks. Our proposed \normalizer offers a promising solution, balancing security and usability by enforcing strict compliance with RFC standards.

\section*{Acknowledgment}

We thank the anonymous reviewers for their valuable feedback. This work was supported by the National Science Foundation under grant 2329540 and was also performed as part of the ARPA-H DIGIHEALS program under Contract No. SP4701-23-C-0089. The views, opinions, and/or findings expressed are those of the author(s) and should not be interpreted as representing the official views or policies of ARPA-H or the U.S. Government.

\bibliographystyle{ieeetr}
\bibliography{main}

\newpage

\appendix

\subsection{Web Framework Parsing Methods}

Table~\ref{tab:framework_parsing} summarizes the methods used for parsing request bodies in our work in each framework. If a content-type is not supported by the default parser of a framework, popular third-party parsers are used instead.

\newpage

\vspace{1.0cm}
\begin{table}[!htbp]
\renewcommand{\arraystretch}{1.25}
\centering
\resizebox{1.0\textwidth}{!}{\begin{minipage}{\textwidth}
\caption{Request Body Parsing Methods in Tested Frameworks}
 \hspace*{1.5cm}
\begin{tabular}{lll}
\toprule
\textbf{Framework} & \textbf{Content-Type} & \textbf{Parsing Method} \\ 
\midrule
Express (Node.js) & application/json & \texttt{request.body} \\ \hline
FastAPI (Python) & application/json & \texttt{request.json()} \\ \hline
Laravel (PHP) & application/json & \texttt{\$request->collect()} \\ \hline
Spring Boot (Java) & application/json & \texttt{Defined a class and used @RequestBody} \\ \hline
Laravel (PHP) & multipart/form-data & \texttt{\$request->collect()} \\ \hline
Express (Busboy) & multipart/form-data & \texttt{form.parse(request)} \\ \hline
FastAPI (Python) & multipart/form-data & \texttt{field1: str = Form("none")} \\ \hline
Flask (Python) & multipart/form-data & \texttt{request.form} \\ \hline
Spring Boot (Java) & multipart/form-data & \texttt{@RequestParam(value = "field1") String field1} \\ \hline
Laravel (PHP) & application/xml & \texttt{\$request->getContent()} \\ \hline
Spring Boot (Java) & application/xml & \texttt{Defined a class and used @RequestBody} \\ \hline
Flask (xmltodict) & application/xml & \texttt{parse(request.data)} \\ \hline
Gin (Go) & application/xml & Defined an \texttt{xmlform} with \texttt{field1} and \texttt{field2} \\ 
\bottomrule
\end{tabular}
\label{tab:framework_parsing}
\end{minipage}}
\end{table}

\end{document}